\documentclass[prb,aps,amssymb,shownopacs,twocolumn,10pt,floatfix]{revtex4-1}

\usepackage{textcomp} 

\usepackage{amsmath}
\usepackage{amsfonts}
\usepackage{amssymb}
\usepackage{mathtools}

\usepackage[utf8]{inputenc}

\usepackage{graphicx}
\usepackage{overpic}

\usepackage{xcolor}

\usepackage{tabularx}

\usepackage{hyperref}

\usepackage{wrapfig}

\newcommand{\specialcell}[2][c]{%
  \begin{tabular}[#1]{@{}c@{}}#2\end{tabular}}

\begin{document}

\title{Detection of Topological Materials with Machine Learning}
\author{Nikolas Claussen$^1$}
\author{B. Andrei Bernevig$^{2}$}
\author{Nicolas Regnault$^{1,2}$}
\affiliation{$^1$ Laboratoire de Physique de l'\'Ecole normale sup\'erieure, ENS, Universit\'e PSL, CNRS, Sorbonne Universit\'e, Universit\'e Paris-Diderot, Sorbonne Paris Cit\'e, Paris, France.\\
$^{2}$Joseph Henry Laboratories and Department of Physics, Princeton University, Princeton, New Jersey 08544, USA}

\date{\today}

\begin{abstract}

Databases compiled using \textit{ab--initio} and symmetry-based calculations now contain tens of thousands of topological insulators and topological semimetals. This makes the application of modern machine learning methods to topological materials possible. Using gradient boosted trees, we show how to construct a machine learning model which can predict the topology of a given existent material with an accuracy of 90\%. Such predictions are orders of magnitude faster than actual \textit{ab initio} calculations. We use machine learning models to probe how different material properties affect topological features. Notably, we observe that topology is mostly determined by the ``coarse--grained'' chemical composition and crystal symmetry and depends little on the particular positions of atoms in the crystal lattice. We identify the sources of our model's errors and we discuss approaches to overcome them.
\end{abstract}

\maketitle


\section{\label{sec:introduction} Introduction} 

Topological insulators (TIs) and topological semimetals (TSMs) are solid state systems which exhibit robust edge or surface modes and quantized bulk response functions due to the topological properties of their electronic wavefunctions \cite{CharlieReview,Bansil2016}. Since the first prediction of TIs in the form of the two--dimensional quantum spin Hall effect \cite{CharlieTI,KaneMeleZ2,AndreiTI}, the field of topological materials has seen remarkable advances. Recently introduced theoretical methods \cite{Bradlyn2017,AshvinIndicators,ChenTCI} have allowed for the large--scale discovery of topological materials.

In particular, the theory of topological quantum chemistry \cite{Bradlyn2017} (TQC) provides a unified framework for the treatment of all topological phases arising from crystalline symmetries. TQC relies on the notion of elementary band representations (EBRs), which enumerate a basis for all electronic bands induced from atomic orbitals (atomic limits), and compatibility relations, constraining how bands can connect across the Brillouin zone.

A set of valence bands which cannot be decomposed as a sum of EBRs makes a material into a strong TI. A TSM is characterized by a set of bands below the Fermi level which does not satisfy the compatibility relations and can therefore not be separated from other bands. While TQC can also fully describe so--called fragile topology \cite{Po2018,Cano2018} (which can be removed by the addition of topologically trivial bands) and polarization topology \cite{multipole} (as occurs in the celebrated Su--Schrieffer--Heeger model \cite{SSH}), in this paper we will focus on strong topology protected by crystal symmetries  and dictated by the symmetry eigenvalues at high-symmetry points in the Brillouin zone. Strong topology is, for the purposes of this paper, defined as the topology that is stable to the addition of any atomic limits.

Applying the theory of TQC to the output of \textit{ab initio} calculations, typically density functional theory \cite{HohenbergKohn,KohnSham} (DFT), on databases of experimentally determined and/or theoretically predicted crystal structures in an automated fashion, large catalogs of topological materials \cite{Vergniory2019,ChenMaterials,AshvinMaterials1,Wieder2019} have been compiled. These efforts showed that far from an isolated phenomenon, topology is ubiquitous: At least 30--40\% of known stoichiometric materials have some nontrivial topological features \cite{Vergniory2019}.

The availability of these datasets, containing tens of thousands of materials, opens the door to modern machine learning (ML) methods \cite{foundations}. The hope is to bypass the complex, multistep computation necessary to determine the topology of a given material by an empirical, statistical model. In particular, such a model could potentially predict the topological features of a material several orders of magnitude faster. Furthermore, a ML model can infer which quantities decide the topology of a material and possibly offer hints on their role, while \textit{ab initio} calculations are very hard to interpret.

In the present paper, we show how to construct a ML model which can predict the DFT--computed topology of a given material with an accuracy of almost 90 \%, based on the dataset of Ref.~\onlinecite{databaseSecret}. Our paper is part of the existing body of research on replacing or accelerating DFT by ML methods \cite{Butler2018,GabrielR.Schlede2019}. Yet, it differs from most of this literature, as we do not search to predict ``energetic'' quantities like the formation energy or the bulk modulus, but an intrinsically quantum property derived from the wavefunction. 

We provide online at \url{https://www.topologicalquantumchemistry.com/mltqc} a fast and efficient tool to predict the possible topological nature of a given material. Beyond the ML model itself, our results include an analysis of various crystal properties with respect to their relevance for the prediction of the electronic topology. We use machine learning models to probe how and how strongly different material properties affect topological features. We find that the model performance saturates at a small number of properties whose impact on the topology we show via a simplified model. In particular, information about the positions of atoms in the crystal lattice does not allow for better predictions, within the limitations of our type of ML model. 

This paper falls under the ML paradigm  known as \textit{supervised learning}\cite{foundations}. The goal is to infer a functional relationship $f: x\mapsto y$ based on a collection of $N$ examples $\{(x_i, y_i=f(x_i))\}_{i=1,...,N}$ which form a so-called sample. A member $i$ of the sample is called a sample point. Among a class of functions, we numerically search for the one which bears the greatest resemblance to $f$ when evaluated on the  sample, as measured by the so--called loss function. This procedure is called training. What sets ML apart from traditional fitting are the large number sample size $N$ and efficient numerical training algorithms, permitting much more complex classes of models. 

This paper is organized as follows: First, we discuss in more detail the dataset on which we rely and the ML methodology which we apply. 
Next, we address the unique challenges which crystal structures pose in the context of ML. We present our model for the prediction of a material's topology and analyze its performance according to different metrics. We use a simplified model to shed light on how the ML model arrives at its predictions and analyze the reasons behind the model's errors. Finally, we highlight obstacles to more sophisticated approaches and discuss our results.

\section{Results}

\subsection{\label{sec:dataset} Dataset}

This paper is based on a large catalog of topological materials which was compiled in Ref.~\onlinecite{databaseSecret} and which is accessible at \url{http://topologicalquantumchemistry.org}. First--principles calculations, using DFT as implemented in the Vienna ab initio Simulation Package \cite{VASP1}, were carried out for 70020 impurity free, crystalline materials selected from the Inorganic Crystal Structure Database (ICSD) \cite{ICSD}. This analysis used nearly $20\time 10^6$  CPU hours. Applying the theory of TQC to the results of the ab initio calculation, materials were grouped into five broad categories \cite{databaseSecret}: insulators with trivial topology, also refered to as Linear Combination of EBRs (LCEBR) (49.5\% of the materials in the database); two types of TIs called Not a Linear Combination of EBRs (NLC) (6.5\%); and Split EBR (SEBR) type (7\%) and two types of TSMs, Enforced Semimetals (ES, 10\% of the dataset) and Enforced Semimetal with Fermi Degeneracy (ESFD, 27\%). The bands of an NLC--type insulator cannot be written as a linear combination of EBRs or parts of an EBR, whereas in SEBRs, there exists an EBR which is energetically split into two bands, of which only one is occupied. A TSM is an ESFD when there is a high-symmetry point degeneracy at the Fermi level and an ES if the degeneracy is away from the high--symmetry points \cite{Vergniory2019}.

Each material is described by its stoichiometric formula, a space group (SG), a unit cell, and the positions of the atoms therein. Many ICSD entries are very similar to each other, describing, for example, different measurements (e.g., at different temperatures) of the same material. All in all, the unique combinations of SG and stoichiometric formula make up only 50\% of the database -- the other materials are ``duplicates''. In 98\% of cases, two materials exhibiting the same stoichiometry and SG also have the same topological classification. Exceptions are mostly due to the instability of DFT calculations (when a small change in the atomic fractional coordinates changes the topological class), or to extreme changes in the environmental conditions (such as ICSD entries describing high pressure measurements). For the purpose of this paper, we therefore group materials into equivalence classes with the same stoichiometry and space group and select only one representative from each group. Thus, the effective size of our dataset is 35009 instead of 70020. 

\begin{figure*}[t]
   \centering
   \begin{overpic}[scale=0.55, angle=-90]{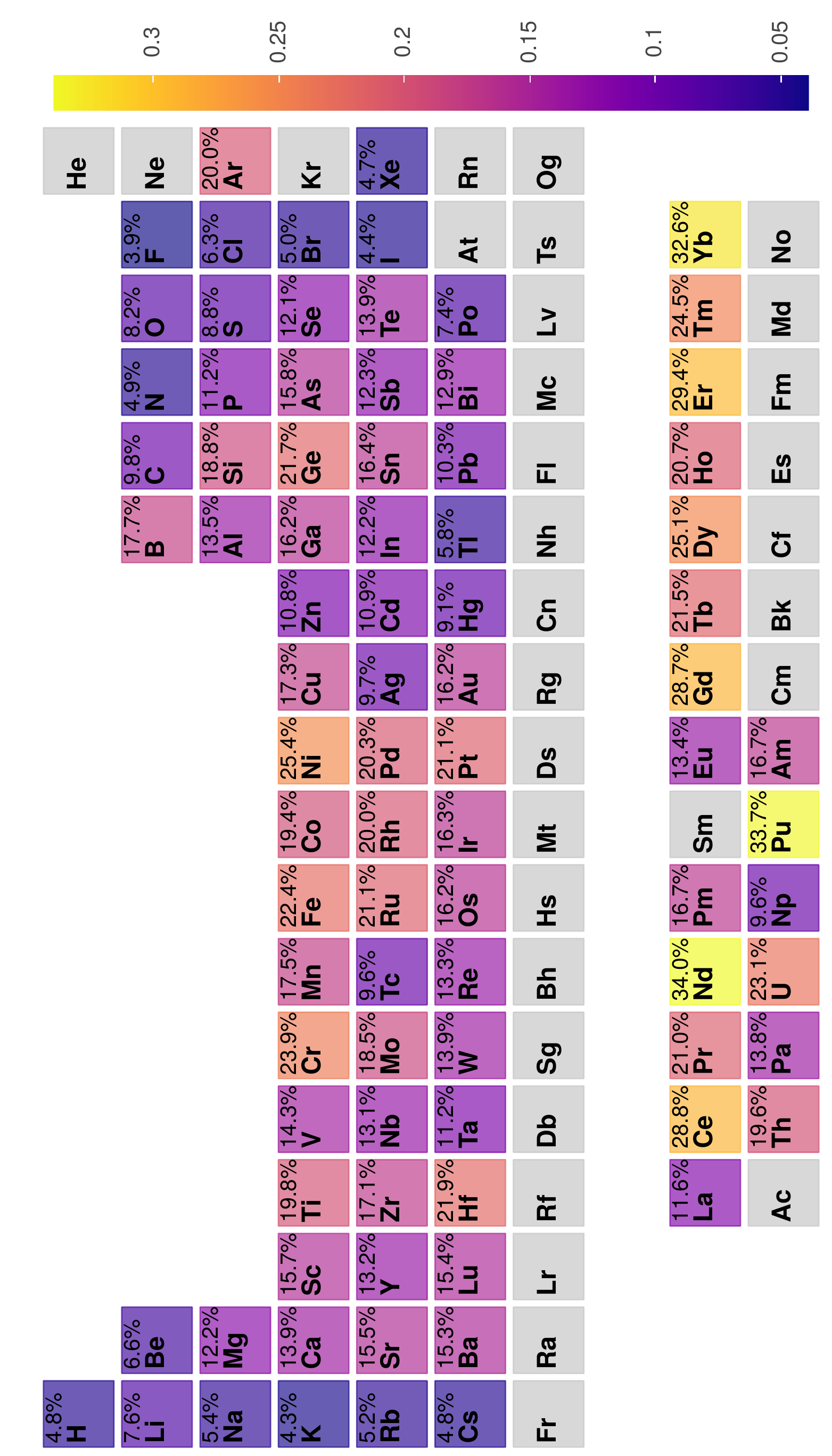}
      \put(102.5,2.5){\rotatebox{90}{\footnotesize $\mathbb{P}[\text{material is TI}|\text{material contains element }x]$}}
   \end{overpic}
   \caption{Chemical composition of TIs (both NLC and SEBR classes). Displayed is the frequency of TIs among compounds containing a given element, which can be interpreted as the conditional probability of a material to be a TI, given that it contains a certain element (the probability is provided by the color scale). This figure shows that not only heavy elements with strong spin--orbit coupling are capable of forming TIs. Elements forming ionic compounds rarely yield TIs, as their conduction and valence band are formed from different atomic orbitals, making band inversion \cite{AndreiTI} unlikely. The alternating pattern among the rare earths could be due to the possibility/impossibility of fully filled $f$--electron bands whenever the number of $f$--electrons is even/odd. However, as $f$--electrons do not typically form dispersive bands \cite{Georges2004}, this is likely an artifact of DFT.}
   \label{figure:periodic_table_TI}
\end{figure*}

The materials in our dataset cover a wide range of complexity, one measure being the number of atoms per primitive unit cell (ranging from 1 to 60), belong to 215 of the 230 different space groups and contain 92 different chemical elements. Fig. \ref{figure:periodic_table_TI} plots the frequency of TIs among compounds containing a given element. Thanks to the large number of materials in the database, we can put heuristics for the chemistry of topological materials  \cite{Bansil2016} on a statistically solid basis. 

\subsection{\label{sec:classification} Classification Models}
\label{MLmethods}

In supervised learning, a classification model is trained by minimizing its prediction error on a sample for which the class of each sample point is known \cite{foundations}. Such a model is often called a \textit{classifier}. We refer to a sample point's class as its \textit{true label}, in contrast to the output of the predictive model, the \textit{predicted label}.

To evaluate a  model after training, we test it by predicting the labels for a second sample or \textit{test set} which was not used during the training process. A detailed description of our testing procedure is given in Appendix \ref{methods:cross_val}. In particular, the scores reported in the body of this paper are computed using tenfold cross
 validation, where the sample is repeatedly (a total of ten times) split into a testing and a training set in different ways. Averaging over the scores eliminates the arbitrariness in the choice of the test set. 

One measure of the quality of predictions on some test set is the accuracy, the fraction of sample points which are classified correctly. When the dataset is imbalanced (i.e. classes are not equally frequent) the accuracy can, however, be misleading. It should be be compared to a naive baseline model which always predicts the most frequent class -- in our case, the baseline accuracy is 49.5\%, i.e. the percentage of trivial insulators. To define additional measures, consider a binary classifier (which distinguishes only two labels, positive/negative). For binary prediction, four cases can occur:

\begin{table}[h]
\begin{tabular}{l|l|l|}
& \specialcell{Predicted label\\positive} & \specialcell{Predicted label\\negative} \\
\hline
\specialcell{True label\\positive} & True Positive & False Negative \\
\specialcell{True label\\negative} &  False Positive & True Negative \\
\end{tabular}
\end{table}

One defines the following quantities:
\begin{align}
\text{Precision} &= \frac{\text{True Positives}}{\text{True Positives }+ \text{False Positives}} \label{eq:precision}\\ 
\text{Recall} &=\frac{\text{True Positives}}{\text{True Positives }+ \text{False Negatives}} \label{eq:recall}\\
F_1 &= 2\cdot\frac{\text{Precision}\cdot \mathrm{Recall} }{\text{Precision} +\text{Recall}}
\label{eq:F1}
\end{align}
\textit{Precision} measures the reliability of a classifier's positive predictions and \textit{recall} measures its ability to find all the true positive sample points. The $\mathrm{F}_1$--score is the harmonic mean of these two quantities. 

If a classifier outputs the probability to belong to a class instead of a mere label, precision can be traded for recall and vice versa by changing the threshold probability after which a sample is classified as positive, yielding the precision--recall curve. When there are more than two classes, there is a precision, recall, and  $F_1$--score for each class, characterizing a model's ability to distinguish a specific class from all others. In this paper, we largely used the $F_1$ score as it provides a single score, largely independent of the choice of threshold, making the comparison between two models straightforward.

\begin{figure*}[t]
	\centering
   \includegraphics[scale=0.35]{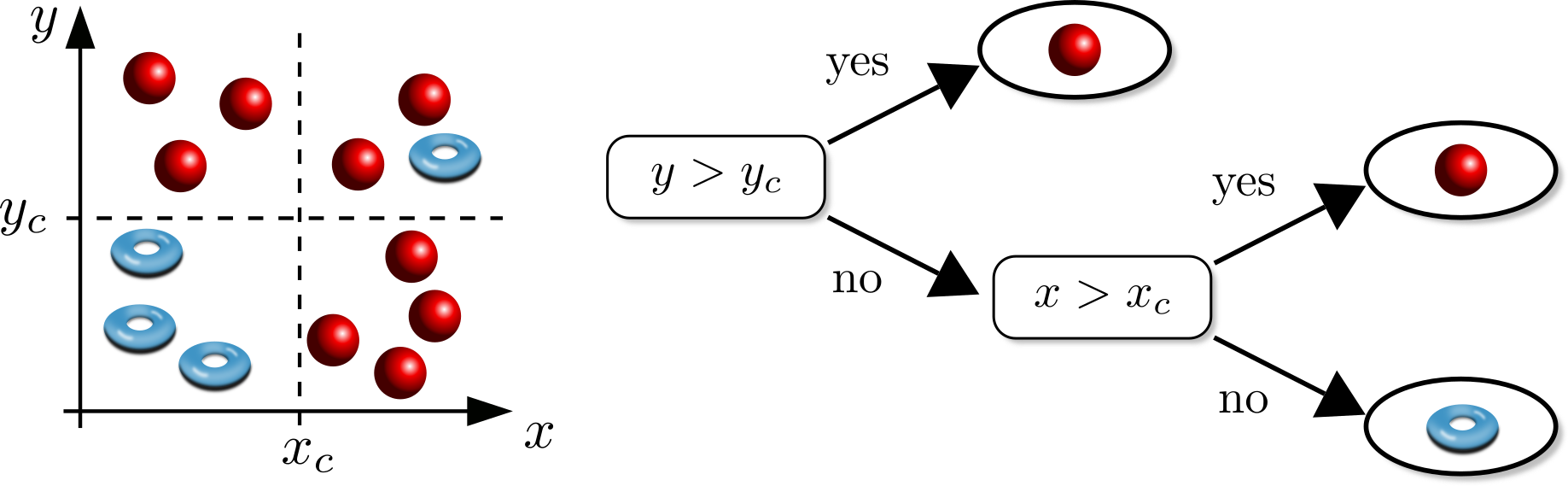}
   \caption{Example of a decision tree. A decision tree recursively splits the dataset according to a set of binary ``questions''. The depth of the tree is the number of sequential questions it poses (here, two questions), and measures its complexity. In the example, a tree with depth 2 cannot perfectly separate tori from spheres but already provides a fairly good classification. Each node of the tree corresponds to a region in the space (here, spanned by $x$ and $y$) of input features, the leftmost node corresponding to the entire space. When constructing a decision tree to classify a sample, the tree is built by recursively adding questions in a left-to-right fashion. At each step, the question which maximizes the information gain about the prediction target at each node is chosen.}
   \label{figure:decision_tree}
\end{figure*}

In this paper, the models  are constructed using the gradient boosted trees (GBT) algorithm  \cite{GB} which is based on decision trees (see Fig. \ref{figure:decision_tree} for an example). In the ML community, GBT is one of the preferred tools for ``tabular'' datasets (those with no underlying spatial or temporal structure, in contrast to images or sounds) such as ours. 
GBT needs no underlying metric to measure the distance between two datapoints (in contrast to a nearest-neighbor classifier, for example) and can therefore naturally deal with situations such as ours where no such metric is available. We empirically tested our GBT model against several other common ML classifiers (random forests, k-nearest neighbor classifiers, linear and Gaussian support vector classifiers \cite{scikit-learn} as well as a fully connected neural network with dropout regularization) and found it to be superior. The best alternative model, a random forest (also based on decision trees), obtains $F_1$--scores of 91, 67 and 90\% for LCEBRs, TIs, and TSMs, respectively (see Table \ref{table:performance}).

GBT is a special case of the \textit{boosting} technique: Instead of training a single, very complex classifier (a so-called strong classifier), boosting linearly combines the predictions of many simple or \textit{weak} ones. Such a combination or \textit{ensemble} \cite{Rokach2010} of weak classifiers is expected to have a lower tendency to overfit \cite{foundations}. In the case of GBT, these classifiers are trees. A weak decision tree is one which has only a small number of nodes. When training a GBT model, one begins by constructing a single tree. Trees are then  added iteratively to the model. Each new tree is constructed to correct the error of the existing model. The technical implementation of the GBT model is described in Appendix \ref{methods:software}.

\subsection{\label{sec:features} Representation of Crystals for ML}

Standard ML algorithms like the decision trees of Sect. \ref{MLmethods} operate on real vectors of a fixed dimension $d$. Each component of such a vector is referred to as a feature. To apply ML models to the classification of crystalline materials, a map  taking crystals to real vectors must be chosen. The choice of this \textit{descriptor} is a non trivial problem the solution of which is crucial for the success of a predictive model. For example, clearly the descriptor must  be independent of the choice of the unit cell and the labeling of a material's constituent atoms. In particular for small sample sizes, the descriptor should also make use of our understanding of chemistry and physics: Elements from the same column in the periodic table are known to be chemically similar. All the while, the dimension $d$ should remain reasonable as the effectiveness of ML models tends to decrease in high dimensions \cite{foundations}. A large number of descriptors for different applications, e.g. organic chemistry, have been described in the literature\cite{MatthiasRupp2011,Bartok2013,K.T.Schuett2014,KatjaHansen2015,MatthewHirn2017,Huo2018,
GabrielR.Schlede2019}. Recent works \cite{Xie2017,KristofT.Schuett2016,JustinGilmer2017} have also proposed neural network architectures tailored specifically to molecules and crystals.

In this paper we adopt an approach which we dub \textit{single--atom statistics}, which has previously been proposed in Ref.~\onlinecite{Ward2016} and used, for instance, in Ref.~\onlinecite{ValentinStanev2018} to predict the critical temperature of superconductors. Single--atom statistics is based on a compound's stoichiometric formula. Each of a crystal's constituent atoms is described by a number of physical and chemical single--atom properties, for example, the number of $s$--shell electrons or the mass. The properties are concatenated to form a vector $\mathbf{X}_i$ for each atom $i$. To derive  the descriptor value for an entire crystal, we calculate statistics over all $N_\text{atom}$  atoms in a unit cell like the mean or the standard deviation:
\begin{align}
\text{Mean}[\mathbf{X}] &:= \frac{1}{N_\text{atom}}\sum_{i\in\text{atoms}} \mathbf{X}_i, \\  
\text{Std}[\mathbf{X}] &:= \frac{1}{N_\text{atom}} \sum_{i\in\text{atoms}} (\mathbf{X}_i-\text{Mean}[\mathbf{X}])^2.
\label{eq:mean_std}
\end{align}
The advantage over a direct encoding of a compound's stoichiometric formula is that we can leverage our knowledge of the physically relevant properties. Let us illustrate this by an example. Elements from the same column in the periodic table are very similar. If we encode the stoichiometric formula directly, replacing an atom in a compound by an atom of a different element of the same column, the encoding vector changes completely. If we use the periodic table column as a descriptor, then the descriptor does not change. Thus, chemically similar substances are represented by nearby vectors, which makes it easier for a machine learning algorithm to infer patterns. Of course, more than one property has to be used, or very different materials might be represented by the same vector.

To validate this approach, we will compare single--atom statistics to a direct encoding of the stoichiometric formula, which we call the \textit{baseline descriptor}.  It consists of a 92-dimensional vector. There are 92 different elements in our dataset, and the $i^\text{th}$ entry equals the fraction of atoms of element $i$ in a compound. For example, NaCl corresponds to the vector $\tfrac{1}{2}\mathbf{e}_{11}+\tfrac{1}{2}\mathbf{e}_{17}$ (Na is element number 11, Cl number 17) and, in general, a compound of the form $X_m Y_n$ corresponds to the vector $\tfrac{m}{m+n}\mathbf{e}_{X}+\tfrac{n}{m+n}\mathbf{e}_{Y}$.

In addition to single atom statistics, we describe crystals by a number of \textit{global} properties like their space group, which pertain to the compound as a whole. Our approach does not take the numeric positions of atoms in the crystal into account.

To decide which single--atom and global properties should be included in our descriptor we began with several hundred physically and chemically motivated quantities, listed in Appendix \ref{methods:features}. We carried out extensive testing to recursively eliminate irrelevant features. At each testing step, we selected one property and temporarily removed it from the descriptor. If the performance (as measured by the $F_1$--scores for the three classes trivial, TI, and TSM) did not drop, the property was permanently deleted from the descriptor, else it was retained. Using this procedure we have extracted those features, discussed below, which are relevant for the prediction of a material's topology. A list of the irrelevant features is given in Table \ref{table:tested_features}. 

The filtering is essential: Indeed, the inclusion of irrelevant features in the descriptor will both decrease the the quality of a predictive model (due to the chance of spurious correlations between an irrelevant feature and the prediction target) and make its interpretation more difficult. We retained a final test set, not used during the filtering, to verify that the filtering did not lead to overfitting (i.e. that the selected properties work only on the sample used to select them -- see Appendix \ref{methods:cross_val}).

The relevant global properties which we have identified are the SG, deciding which type of topology protected by crystalline symmetry is possible \cite{Bradlyn2017}, and the number of electrons per unit cell ($N_e$), deciding whether bands in certain space groups are necessarily completely filled or not. For example, in time--reversal symmetric crystals, the Kramers degeneracy implies that a compound with odd $N_e$ is necessarily a semimetal \cite{Vergniory2019}. Appendix \ref{methods:categorical} describes in detail how these properties enter the descriptor map. Among the single--atom properties, the most relevant ones are the mean number of $s,p,d,f$--shell valence electrons and the mean and standard deviation (as defined in Eq. \ref{eq:mean_std}) of the atom's row and column number in the periodic table. For example, NaCl has one atom from columns 1 and 31 each, giving a mean column number of 16 and a standard deviation of 15. We will refer to this collection of eight features (mean number of $s,p,d,f$, mean and variance of column and row number) as $spdf+$. Finally, the number of atoms in a compound from each column and row in the periodic table is relevant. As an example, half of the atoms in NaCl are in column 1 and 32 respectively, so the column fractions would be represented by the vector $\tfrac{1}{2}\mathbf{e}_1+\tfrac{1}{2}\mathbf{e}_{32}$, where $\mathbf{e}_i$ is the $i^\text{th}$ unit vector. 

To include the numeric positions of atoms in a descriptor we used a technique described in Refs.~\onlinecite{Ward2017,OlexandrIsayev2017,Ward2018}. It calculates differences in single--atom quantities between nearest neighbors, weighted by the boundary area shared by two neighbors (computed using the Voronoi tessellation\cite{Ward2018}). An example is the mean difference in electronegativity of neighboring atoms. However, none of these features, to which we refer as ``nearest--neighbor differences'', proved relevant to our classification problem, as shown by Table \ref{table:performance}. This finding is consistent with the fact that two ICSD entries with the same stoichiometric formula and SG, but possibly differing atom positions, almost always have the same topological class. 

\subsection{\label{sec:performance} Performance of the GBT Model} 

With the features described in Sect. \ref{sec:features} and the ML models of Sect. \ref{sec:classification} we are able to build a classifier which can predict the topology of a given material with an  accuracy of 90\%, compared to a baseline of 49.5\%. 

\begin{figure*}[t]
    \centering
   \begin{overpic}[scale=0.3]{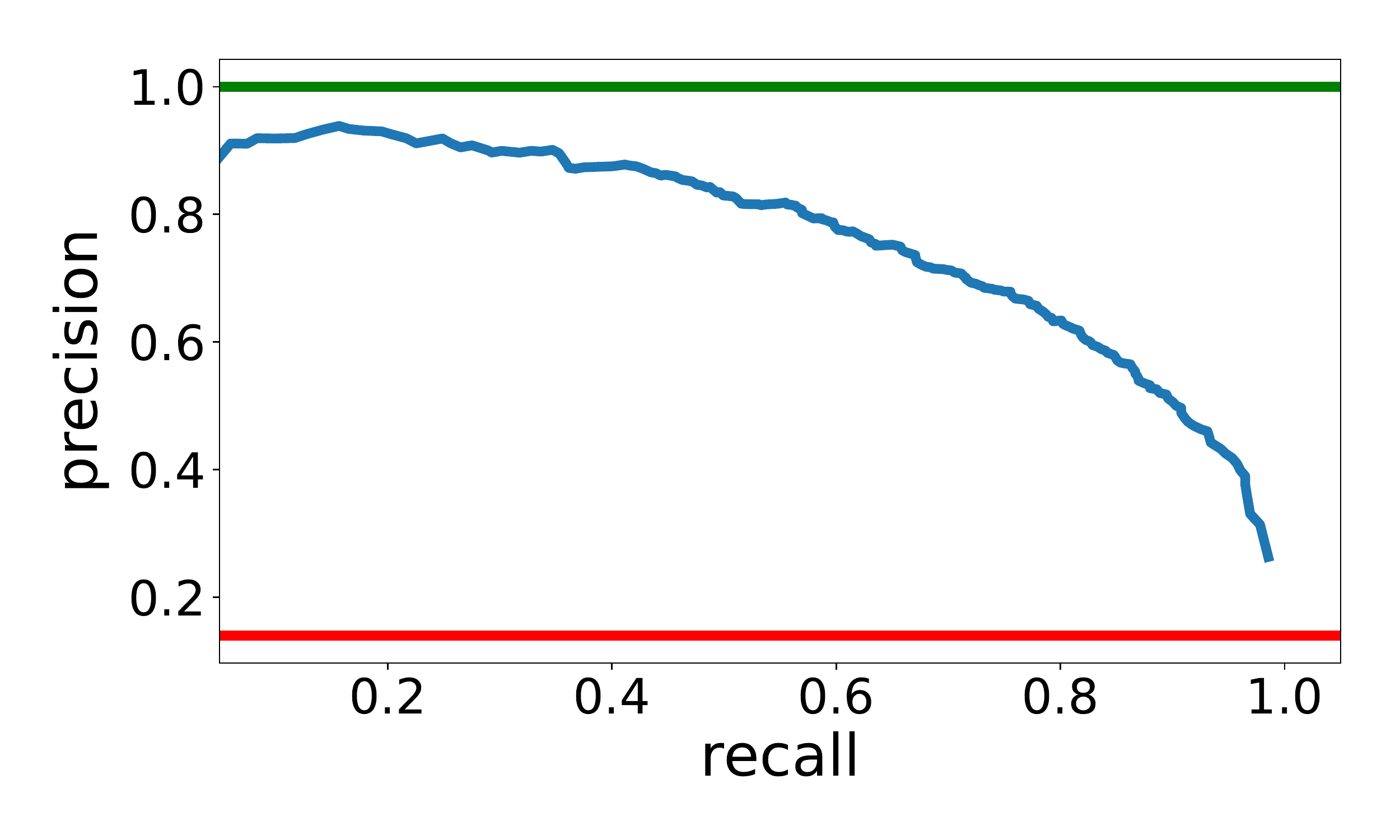}
      \put(2.5,52.5){\large \textbf{a}}
   \end{overpic}
   \begin{overpic}[scale=0.3]{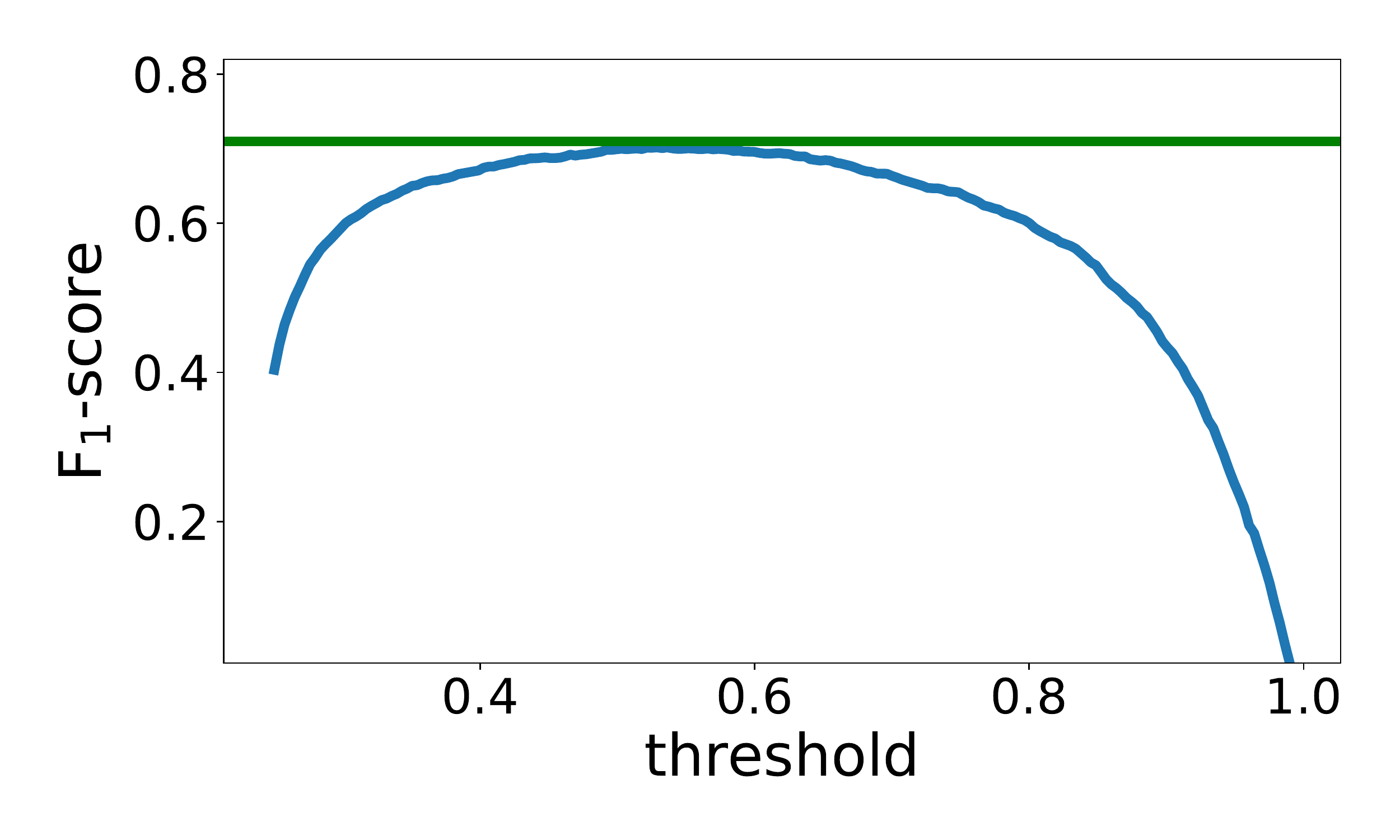}
      \put(2.5,52.5){\large \textbf{b}}
   \end{overpic}      
    \caption{\textbf{a} Precision--recall curve \ref{table:performance} and \textbf{b} $\mathrm{F}_1$--score as a function of the classification threshold, both for the full GBT model from Table \ref{table:performance}. We focus here on the detection of TIs, the topological class which is both the most interesting and the hardest to detect. Green lines indicate the performance of a perfect model, where high precision and high recall can be achieved at the same time (left), respectively, where precision and recall can be tuned without an overall loss of performance (right). The red line in the left panel shows the null model, which randomly assigns labels according to the threshold probability (the precision is then equal to the overall frequency of TIs in the dataset; see Appendix \ref{methods:random_classifier}). These plots show that over a wide range recall and precision can be exchanged without affecting the overall performance as measured by the $F_1$--score. A model tuned to have a high precision could, for example, help find promising candidate materials in a database, to be verified by DFT. The left graph shows that at a precision of 90\% we have 30\% recall.}
   \label{figure:precision_recall_curve}
\end{figure*}

This is demonstrated by the first line of Table \ref{table:performance}, showing the $F_1$--score for the different topological classes, as well as the precision--recall curve of Fig. \ref{figure:precision_recall_curve}. Notably, the algorithm performs remarkably well on TSMs and on trivial insulators, while being less efficient for TIs. We attribute this difference to both the much lower frequency of TIs in the dataset and the more subtle nature of their topology. For instance, whether a material has inverted bands is more difficult to infer from the stoichiometric formula than whether that material has a large gap due to ionic bonds.

\begin{table*}[t]
\begin{tabularx}{\textwidth}{|l|X|l|l|l|l|l|}
  \hline
  Model & Descriptor & $d$ & Acc.  & $F_1$ Triv. & $F_1$ TI & $F_1$ TSM \\
	& & & [\%] & [\%] & [\%] & [\%] \\	  
  \hline
  Full model (FM) & SG, $N_e$, $spdf+$, number of atoms from each periodic table row and column & 49 & 89.7(5) & 94.0(3) & 70(1) & 92.0(5)  \\ 
  \hline
  FM + Non-SOC & features used by FM and topological classification of material obtained by DFT without SOC & 50 & 92.0(3) & 96.5(2) & 77(1) & 93.3(4)  \\ 
  \hline
  Baseline model & SG, $N_e$, baseline descriptor (number of atoms from each element in the stoichiometric formula) & 94 & 86.0(5) & 92.5(5) & 67(1) & 91.0(5) \\ 
  \hline  
  $spdf+$ model & SG, $N_e$, $spdf+$ features & 10 & 87.7(5) & 93.0(5) & 69(1) & 92.0(5) \\ 
  \hline 
  FM + nearest neighbor & features used by FM and nearest--neighbor difference features \cite{Ward2017}, defined in Sect. \ref{sec:features} & 184 & 89.0(5) & 94.0(3) & 69(3) & 92.0(5) \\ 
  \hline
  FM without SG & $N_e$, $spdf+$, number of atoms from each periodic table row, and column & 48 & 84.0(5) & 91.5(3) & 57(2) & 86(1)  \\ 
  \hline 
\end{tabularx}
\caption{\label{table:performance} Performance of GBT models as measured by the $F_1$--score and accuracy (Acc.). The accuracy should be compared to the baseline of 49.5\%. We display the mean and in brackets the standard deviation of tenfold cross--validated scores (see Appendix \ref{methods:cross_val}). The models differ by the descriptor used to represent materials, defined in column 2. Details on the encoding of $N_e$ and SG are given in Appendix \ref{methods:categorical}. $d$ is the total number of properties included in the descriptor.}
\end{table*}

We can also configure the model to predict the topological subclass (NLC, SEBR, ES, and ESFD). Table \ref{table:subclasses} shows the resulting $F_1$--scores. Overall, the scores are lower: With more classes come lower sample sizes for each class and more opportunities for misclassification, and thus lower performance. For example, a NLC being classified as a SEBR would not count as error when simply predicting whether a material is a TI.

Some conclusions can be drawn. ES semimetals are harder to detect than ESFD semimetals because the latter can in many cases be detected from $N_e$ and the SG alone. We see that our model reproduces these rules. Likewise, our model correctly identifies the groups in which TIs are allowed according to TQC. Regarding the TIs, NLCs are easier to predict than SEBRs. We attribute this to the more pronounced dependence of SEBR--type topology on energetics, which is not fully captured by our model.  

\begin{table*}[t]
\begin{tabular}{|l|l|l|l|l|l|}
  \hline
   Total accuracy [\%] & $F_1$ Triv. [\%] & $F_1$ NLC [\%] & $F_1$ SEBR [\%] & $F_1$ ES [\%] & $F_1$ ESFD [\%]  \\
  \hline
  \hline
   87.0(3) & 94.0(4) & 66(2) & 59(3) & 73(2) & 95.5(3) \\
  \hline
\end{tabular}
\caption{Performance of the FM when predicting the topological sub--class, as measured by the $F_1$--score and accuracy. The accuracy should be compared to the baseline of 49.5\%.}
\label{table:subclasses}
\end{table*}

Additionally, we can include partial information from \textit{ab initio} calculations in the model. As most TIs are the result of band inversion driven by spin--orbit coupling (SOC)  \cite{Bansil2016}, DFT must be performed with SOC to detect topological materials. However, as Table \ref{table:performance} shows, already the results of much less expensive calculations without SOC help our model significantly. 

\subsection{\label{sec:analysis} Analysis of the GBT Model}

Let us first comment on the features chosen by our filtering procedure to describe a crystal. Table \ref{table:performance} shows that a model that only uses the mean number of $s,p,d,f$ valence electrons, the mean and variance of the periodic table column and row as well as SG and electron number can capture most of the performance of our full model. It also outperforms the baseline model which directly translates the chemical formula into a vector. This validates the single--atom statistics approach (see Sect. \ref{sec:features}).

The important role of global properties like the SG is shown by the pronounced drop in performance when these features are left out. Finally, we also included much richer collections of properties, the nearest--neighbor differences defined in Sect. \ref{sec:features}, which take into account the positions of atoms in the unit cell. Table \ref{table:performance} shows that they actually perform slightly worse as the inclusion of irrelevant features in a descriptor increases the risk of overfitting.

Let us emphasize that when a feature does not increase the performance of our predictive mode, we cannot conclude that it is irrelevant for a material's topology -- this could simply be due to the failure of the particular ML approach. Yet, the model performance without a certain feature gives an upper bound on the potential feature's influence on the prediction target. 
For instance, suppose the accuracy of a model on some sample is $a\%$. We can then conclude that the features \textit{not} available to the model only influence the label or class of $(100-a)\%$ of the sample.

The other way around, the relevance of some quantity for the model does not imply that it is physically decisive. Rather, it could be only correlated with a physically relevant quantity. However, we know that the input of our model is causally related to the topological class (for instance, TQC shows how changing a material's SG by breaking a crystal symmetry will change the topology), ruling out a mere correlation.

Therefore, we believe that we can draw the following conclusions. Given the space group, the positions of a compound's atoms within the crystal lattice are of limited importance for the topology of a material. Rather, it is the ``average orbital character'', the mean number of $s,p,d,f$--shell valence electrons which counts. Furthermore, we showed the central importance of a material's SG. This hints towards the degrees of freedom which a minimal phenomenological model should take into account.

\begin{figure*}[t]
	\centering 
   \includegraphics[ trim={1cm 1cm 0 1cm}, scale=0.53, clip]{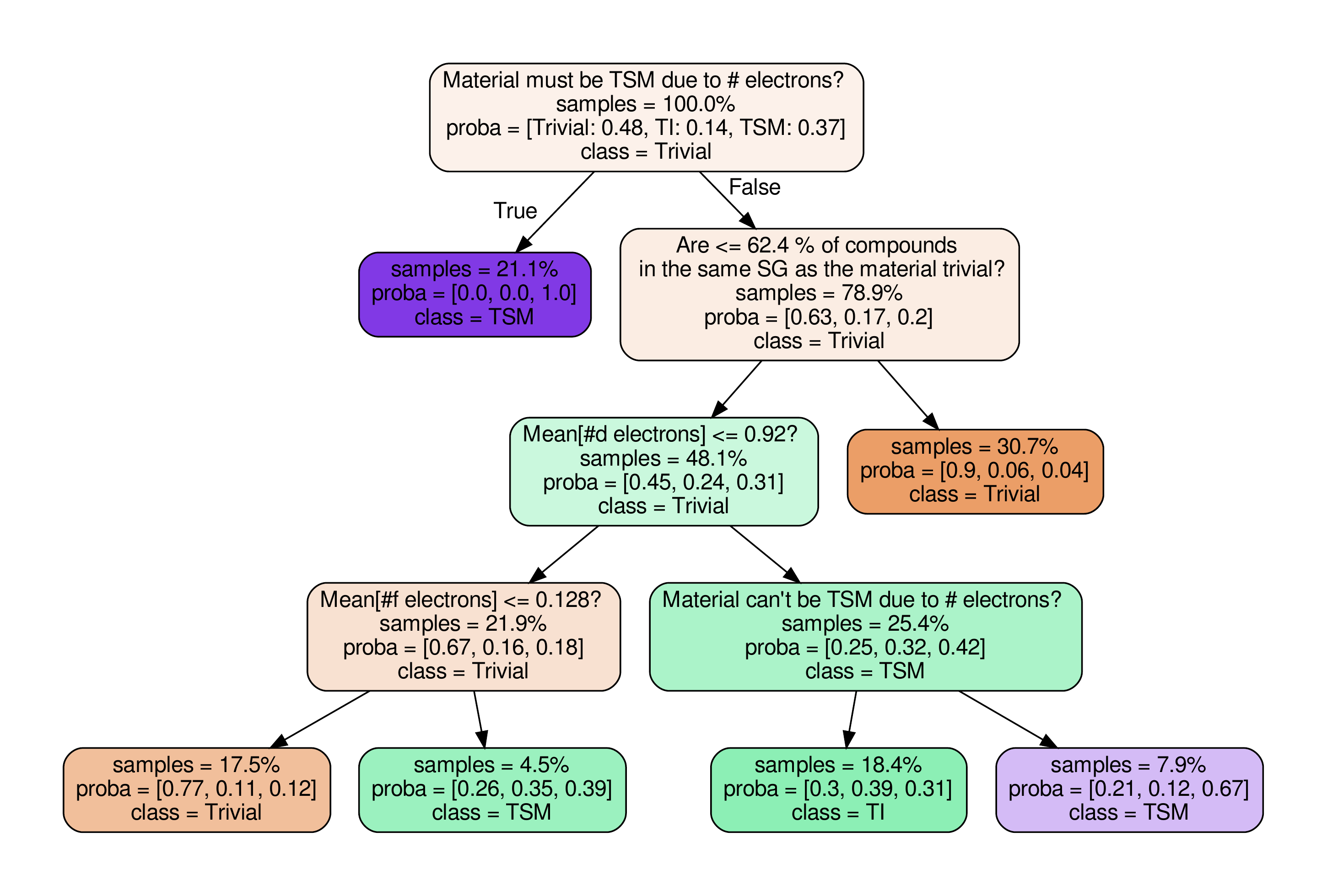}
   \caption{Simplified tree model. The descriptor used is that of the $spdf+$ model (see Table \ref{table:performance}). To predict the topology of a given material, these features are first calculated. The tree then asks a series of yes/no questions. To understand the graph, consider the top node. Recall from Fig. \ref{figure:decision_tree} that each node of the tree corresponds to a subset of the dataset, the top node corresponding to the entire dataset. The ``proba'' field gives the probability to belong to class trivial/TI/TSM and the ``class'' field states the most probable one. At the top node, the probabilities are equal to the frequencies of the classes in the entire dataset and the ``samples'' field, which gives the proportion of the dataset which arrives at a node, is equal to 100\%. 
    The top node asks whether the material has to be a TSM due to its SG and $N_e$. Depending on the answer, a material is either classified as TSM or further questions are asked. The next question for instance considers the SG and asks whether the material belongs to a SG the members of which are mostly trivial insulators.
	While our full model has no input on which combinations of SG and $N_e$ lead to TSMs and learned those rules by itself, for this simplified tree we computed the answer to the question ``Material must be TSM due to no. of electrons?'' and `Material cannot be TSM due to no. of electrons?'' for each material in advance and used it as an input feature. This allows us to keep the tree at a readable size: the tree does not need to ask numerous questions on $N_e$ and SG to answer these questions by itself.}
   \label{figure:simple_tree}
\end{figure*}

Next, we want to understand how our model makes use of these features. ML models are often treated as ``black boxes'': the complexity of their inner workings which makes their high performance possible obfuscates how they arrive at predictions. In the case of a GBT model, the large number of trees which make up the model obstructs its analysis. To address this, we have developed a simplified model which uses only a single tree, depicted in Fig. \ref{figure:simple_tree}. Despite its simplicity, it performs surprisingly well, with $F_1$--scores of 84\% (trivial), 48\% (TI) and 80\% (TSM). The tree illustrates explicitly how the GBT model can arrive at its decisions and provides quantitatively supported heuristics for finding topological insulators.

We can also obtain some interesting physical insight from this simplified tree. It shows that a large number of $d$-- or $f$--shell electrons help to turn a material into a TI. A necessary precondition is a SG which not only theoretically permits but favors (crystalline) topological over trivial insulators. Typical examples are many SGs within the point group $D_{2h}$ (and to a lesser degree, $D_{4h}$), for instance, SG 55. As expected, we recover that TSMs are mostly determined by $N_e$ and SG (due to the ``enforced semi--metal'' property \cite{Bradlyn2017}). See the supplementary material of Ref.~\onlinecite{Vergniory2019} for a table indicating which combinations of SG and $N_e$ lead to a TSM.

\subsection{\label{sec:limitations} Limitations of the GBT Model} 

\begin{figure*}[t]
	\centering
   \includegraphics[scale=0.3]{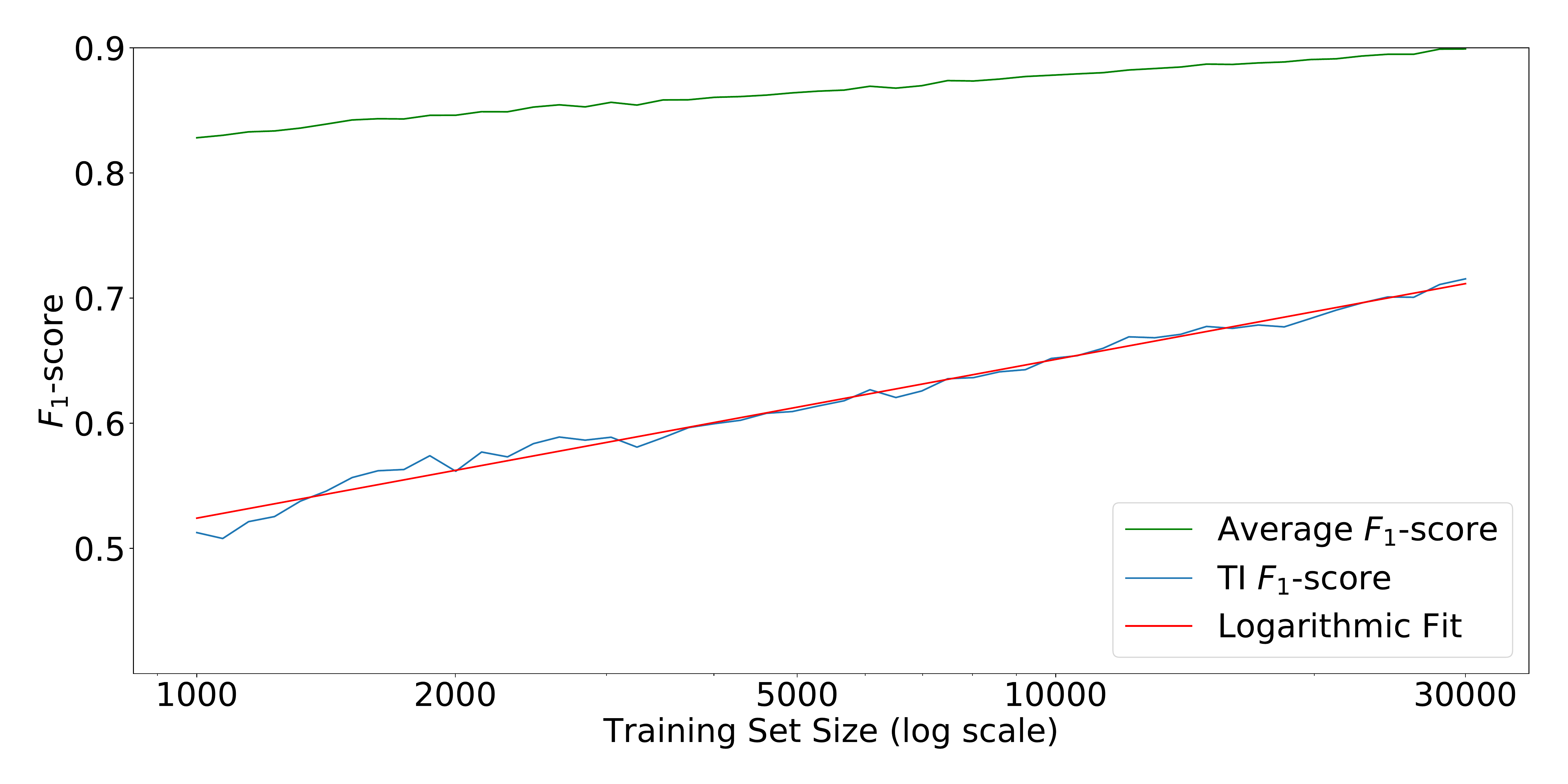}
   \caption{Dependence of $F_1$--score on the training set size. The average $F_1$--score is the average of the $F_1$--score for the three classes. Performance increases logarithmically with the training set size. A doubling of the training set size increases the $F_1$--score for TIs by approximately 0.05. Note that a $F_1$--score is always $\leq 1$, so the $\log$ dependency cannot hold when the $F_1$--scores approach this upper bound.}
   \label{figure:training_set_size}
\end{figure*}

The main limitation of any ML model is the number of samples available for training. We show in Fig. \ref{figure:training_set_size} how the performance of the model scales with the training set size. Our dataset contains a total of approximately 2500 SEBR-- and NLC--type TIs each. This modest sample size is a crucial roadblock. In addition, the materials in the database are very diverse, belonging to many different SGs and containing many different chemical elements. Therefore, for a given topological material, there are often only a few other similar ones, making it harder to infer patterns.

Furthermore, we know that the DFT calculations used to determine the topological labels on which we rely can be unreliable. In such cases, the label may be somewhat random and therefore essentially impossible to predict. Based on a number of criteria (detailed in Appendix \ref{methods:nice}), we selected a subset of 4009 materials for which we are rather confident of the DFT predictions. We will refer to them as \textit{high--confidence} materials. In particular, we selected TIs with high gaps and excluded magnetic and $f$ electron materials. The results are displayed in Table \ref{table:subclasses_nice}. With respect to Table \ref{table:subclasses}, we see significant improvements, which supports the hypothesis that some of the model's errors are due to randomness in the DFT--calculated labels. From a point of view of the search for real--world (not just DFT--calculated) topological materials, it is encouraging to see that large--gap TIs are correctly identified more often. 

\begin{table*}[t]
\begin{tabular}{|l|l|l|l|l|l|}
  \hline
   Total accuracy [\%] & $F_1$ Triv. [\%] & $F_1$ NLC [\%] & $F_1$ SEBR [\%] & $F_1$ ES [\%] & $F_1$ ESFD [\%]  \\
  \hline
  \hline
   92.0(5) & 96.0(4) & 72(3) & 68(1) & 79(1) & 99.0(5) \\
  \hline
\end{tabular}
\caption{Performance of the FM evaluated exclusively on high--confidence materials. Here, we predict the topological sub--class and measured the $F_1$--scores and accuracy. Details on the selection and testing procedure are given in Appendix \ref{methods:nice}.}
\label{table:subclasses_nice}
\end{table*}

\begin{figure*}[t]
	\centering
   \includegraphics[scale=0.3]{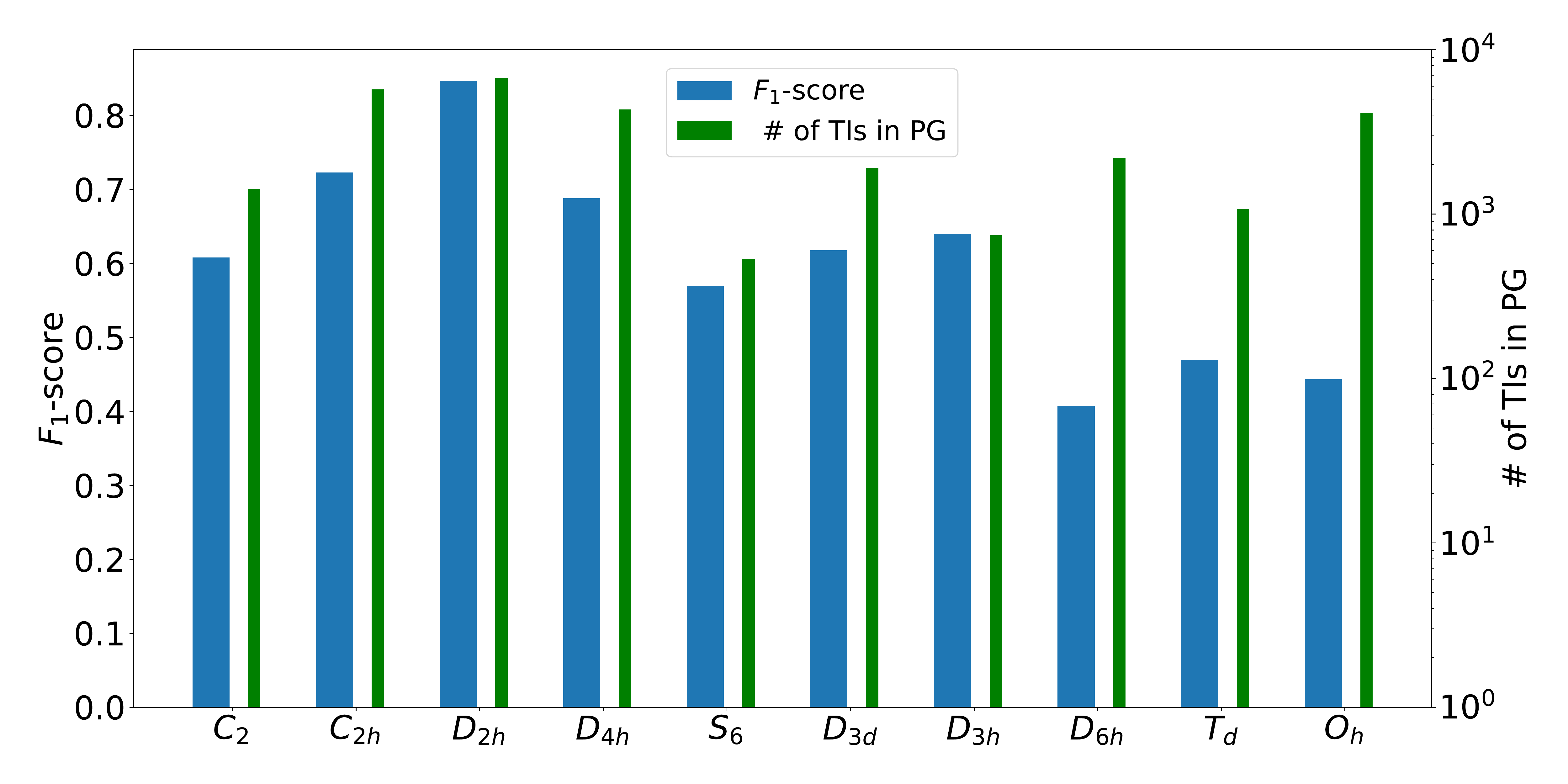}
   \caption{Dependence of the $F_1$--score for TIs on the point group (PG). The scores are calculated by testing the model on subsets of the entire test set which contain only materials from a single point group. Here, we use the more coarse--grained PG instead of the SG to obtain larger subsets. Note that the average of the $F_1$--scores shown in this plot is not equal to the $F_1$--score on the entire test set: the different subsets contain different proportions of false and correct predictions and the $F_1$--score is not a linear function of these arguments (see Eq. \ref{eq:F1}). Only PGs which contain any TIs are shown. These are all PGs with an improper rotation (some are not included in the plot, as they do not occur in the dataset frequently enough), a pattern which is correctly learned by the model. Note that for materials from certain PGs (namely, the cubic $\bar{4}3m$ and $m\bar{3}m$ and the hexagonal $6/mmm$) we observe much lower performance. We do not have a clear explanation for this discrepancy. We checked that it cannot be solely explained by different ratios of NLC-- to SEBR--type TIs (the latter are known to be harder to detect) or the differing frequency of TIs in different PGs (see Appendix \ref{methods:random_classifier}).}
\label{figure:point_groups}
\end{figure*}

We can also pinpoint specific types of materials where our model performs poorly. Fig. \ref{figure:point_groups} shows that the error rate of the model strongly depends on the symmetry properties of the materials: For cubic point groups, as well as for the hexagonal point group $D_{6h}$, the model seems to detect topological insulators much less accurately. This is consistent with the fact that these point groups are more complicated in the sense that they contain the greatest number of distinct symmetry operations of all point groups \cite{BigBook}. 

The error rate also depends strongly on the chemical composition: For materials which contain alkali metals or halogens, the recall for TI is lower than average. This is unsurprising, as the ionic crystals typically formed by these elements are not typically TIs (see Fig. \ref{figure:periodic_table_TI}). 

\section{\label{sec:discussion} Discussion} 

In Ref.~\onlinecite{databaseSecret}, the authors compiled the largest--yet catalog of topological materials. Based on this work, we have investigated the prediction of topological classes with machine learning. We showed that a simple and robust ML model, based on gradient boosted trees, can predict the topology of a material's electronic structure with high accuracy, using only its chemical composition and symmetry properties, without costly DFT calculations. Since our ML model cannot guarantee that a real material has topological features, our approach does not claim to replace a full fledged DFT calculation. Still, it provides a fast and efficient tool to predict the possible topological nature of a given material. 
We also caution that a ML model is only as good as the dataset it was trained on: In spite of its success \cite{Vergniory2019,ChenMaterials,AshvinMaterials1,Wieder2019}, there is no theoretical guarantee that DFT always correctly captures the topological class, which is derived from the electronic wave function. DFT however principally models functions of the electron density. We address this issue in Appendix \ref{methods:nice} by considering a subset of materials where we are highly confident in the DFT predictions. Our ML model and the code used to construct it are available online at \url{https://www.topologicalquantumchemistry.com/mltqc}. 

The main result of this paper -- beyond the predictive model itself -- is the use of ML models to probe the relevance of different properties of materials to their topology. We find that our algorithm performance saturates at a very small number of input features, which we identified.
Notably, we observe that topology is mostly determined by ``coarse--grained'' chemical composition and crystal symmetry and depends little on the particular positions of atoms in the crystal lattice. We constructed a simplified model, where the impact of these features can be seen directly. In particular, materials with a large number of $d$-- or $f$--shell valence electrons in general, and not only compounds containing heavy elements with strong SOC, are likely to be topological insulators.

We investigated several approaches to overcome the limitations and improve the predictions of our model by including the atomic positions in our model, but without success, for instance, the following. TQC emphasizes the decisive role of the symmetry properties of the particular point in the crystal lattice on which an orbital is centered, its \textit{Wyckoff position}. Orbitals on certain Wyckoff positions will induce split EBRs (and thus give rise to SEBR--type TIs). We refer to these positions as SEBR--inducing Wyckoff positions (SEBRiWPs). It seems natural and has been conjectured \cite{Cano2018} that compounds with atoms on SEBRiWPs  will likely have topological bands. Note that while this does not necessarily imply topology at the Fermi level because these bands can have any energy, the existence of topological bands does make topology at the Fermi level more likely. However, we find that only 46\% of SEBR--type materials have atoms on SEBRiWPs. This number is not significantly higher than for the materials from other topological classes. Unsurprisingly, the information on whether a material has atoms on SEBRiWPs does not help our predictive model. To make sense of this, let us note that an SEBR is not obliged to have atoms at some SEBRiWP -- it is only adiabatically equivalent to a crystal which does. Our results suggest that the SEBR wave functions obtained by \textit{ab initio} calculations are perhaps very different from the idealized ones to which they are adiabatically equivalent. This makes it difficult to use the atomic positions in a predictive model.

Future research should try to take into account the spatial structure of crystals. Given the obstacles we encountered, we believe in the need for more sophisticated ML architectures (such as crystal graph convolutional networks (CGNNs) \cite{JustinGilmer2017}). Such architectures can operate directly on crystals and allow one to use the numeric positions of atoms without the need for an explicit descriptor (see Appendix \ref{methods:features}). However, CGNNs suppose that the quantity to be predicted can sensibly be written as a sum of local contributions, which is appropriate for, e.g., the enthalpy of atomization but not necessarily for predicting the topological class. A challenge is therefore the inclusion of global quantities like the crystal symmetries into these architectures. Alternatively, we must leverage physical understanding to ``pre--process'' the data on a crystal's spatial structure, making it easier for an ML algorithm to infer patterns. One promising candidate is the use of the empty lattice approximation \cite{AshcroftMermin}, which can allow an estimate of topological features from the lattice constants only.

\begin{acknowledgments}
We cordially thank L. Elcoro for his support by providing data from the Bilbao crystallographic sever. We also thank S. Biermann and M. G. Vergniory for helpful remarks and discussions. B.A.B. and N.R. are supported by the Department of Energy Grant No. desc-0016239, the National Science Foundation EAGER Grant No. DMR 1643312, Simons Investigator Grants No. 404513, No. ONR N00014-14-1-0330, No. NSF-MRSECDMR DMR 1420541, the Packard Foundation No. 2016-65128 and the Schmidt Fund for Development of Majorama Fermions funded by the Eric and Wendy Schmidt Transformative Technology Fund.
\end{acknowledgments}

\textbf{Note added:} Recently, its was brought to our attention that a related work \cite{Andrejevic2020} has appeared.

\appendix

\section{\label{methods:cross_val} Cross Validation and Test Set}

Cross--validation allows one to eliminate the dependence of the performance estimate on the way the sample is split into the testing and the training set \cite{foundations}. The entire sample, of size $N$, is randomly partitioned into $k$ subsets of size $N/k$, called folds. In stratified cross--validation one additionally ensures that each fold contains the same proportion of sample points from each class as the entire sample. Now, for each $i\in \{1,...,k\}$ the model is trained using the data from folds $1,...,i-1,i+1,..k$ and tested on fold $i$. One can then average over the $k$ scores from the $k$ folds (the fold scores) to get a better performance estimate. We use tenfold stratified cross--validation for the estimates stated in Tables \ref{table:performance} and \ref{table:subclasses}. In cross--validation the test size is restricted to be equal to $N/k$ for some $k$, so Fig. \ref{figure:training_set_size}, which displays the model performance as a function of the training set size, was calculated by an average over five random splits into test and training set for each point on the graph. 

At the beginning of the ML project, we split the dataset into two parts. We used 32179 materials to develop our GBT model (by analyzing the data and testing different models), and the remaining 2830 materials were used only to test the final model. We refer to them as the final test set. This practice safeguards against overestimating the performance we can achieve by choosing one among several models under investigation (in our case, mostly GBT models using different features as well as some other ML models investigated before we decided to use GBT).

This can be understood as follows. A collection of many different models $i=1,...,n$ can be thought of as a single ``meta--model'', in which the index $i$, i.e., which model to use, is an additional parameter. Optimizing $i$ by choosing the model which performs best on a sample is then a training procedure precisely analogous to optimizing the parameters of a single model. Therefore, overfitting as described in Sect. \ref{sec:classification} can also occur and an independent final test set is necessary to faithfully evaluate  the metamodel performance.

However, because of the limited number of materials in the database and the strong variability of the model error between different materials as discussed in Sect. \ref{sec:limitations}, our final test set is too small to give accurate estimates of the model performance. This is evidenced by the standard deviations of the cross--validated scores given in Table \ref{table:performance}. We instead checked that the performance on the final test set is within a standard deviation of the mean cross--validated score on the entire dataset of 35009 materials. For our full model, we find $F_1$--scores of trivial, 94\%; TI, 69\%; and TSM, 92\%, on the final test set. In spite of minor differences, this agrees to within a standard deviation with the cross--validated scores shown in Table \ref{table:performance}, showing that we are are not overestimating performance due to model selection.

\section{\label{methods:software} Technical Implementation}

This paper uses the \texttt{sklearn} \cite{scikit-learn} framework for statistical learning for the computation of cross-validated scores and for the training of the single--tree model. The main models are trained using the \texttt{xgboost} \cite{XGB} implementation of gradient boosting. Certain features, for instance, the nearest--neighbor differences, were calculated using \texttt{matminer} \cite{Ward2018}.  The code written for this paper will be made available online at \url{https://www.topologicalquantumchemistry.com/mltqc}.

\subsection{\label{methods:hyper} Hyperparameters}

The GBT algorithm controls the complexity of the constructed model with several hyperparameters like the maximal tree depth. Increasing the model complexity increases the tendency to overfit. For instance, in a tree of depth $\log_2(N)$ (in our case, $\log_2(N)\approx 15$), the leaf number (the number of terminal nodes in a tree; in Fig. \ref{figure:decision_tree} there are three) equals the sample size $N$, so that the tree can simply memorize the labels of all training sample points. This will not generalize to new data. If the model complexity is too low, the model cannot fit the data well. 

The hyperparameters were chosen through cross--validated random search in the parameter space: We trained models for 300 different randomly chosen values of the hyperparameters, evaluated their performance by cross--validation, and chose the model with the best $F_1$--score. Because there are three different $F_1$--scores for each model, the ``best'' model is not uniquely defined, but we found that the three scores are strongly correlated when hyperparameters are varied. So in practice, this problem was not very pressing. 

The parameters we chose are listed in Table \ref{table:hyper}. The same parameters were used to compute all scores given in this paper (in order to ensure comparability between models using different descriptors). Of particular importance are the maximal tree depth and the minimal child weight, which controls the minimal number of samples which belong to a tree's leaf (a low value is necessary to deal with SGs and elements which are very rare). Column subsampling randomly selects a given percentage of all features and, when training a new tree, increasing robustness to overfitting. Learning rate and $L^2$ regularization regularize the weights assigned to individual trees. We furthermore used a weighted loss function to offset the class imbalance. The full model of Table \ref{table:performance} contains 150 individual trees. Recall that during the training of a GBT model trees are added to the model one after the other. We computed the performance after each such step and found that it saturates after $\sim 150$.  

\begin{table*}[t]
\begin{tabular}{|l|l||l|l|}
  \hline
   Parameter & Value & Parameter & Value \\
  \hline
  \hline
   Maximal tree depth & 10 & Minimal child weight & 0.1  \\
  \hline
  Learning rate $\eta$ & 0.23  & $L^2$ regularization $\lambda$ & 1.33 \\
  \hline
   Column subsampling by tree & 0.78 & Column subsampling by node & 0.75 \\
  \hline
\end{tabular}
\caption{Hyperparameters for the full GBT model. A full definition of all parameters can be found in Ref.~\onlinecite{XGB}.}
\label{table:hyper}
\end{table*}

\section{\label{methods:features} Tested Features and Unsuccessful Models}

\begin{table*}[t]
\begin{tabularx}{\textwidth}{|l|X|}
  \hline
   Category & Tested Features \\
  \hline
  \hline
  Features derived from SG &  Bravais lattice, point group, subgroups of the space group \\
  \hline
  Unit Cell & lattice parameters, their ratios, angles of lattice vectors, unit cell volume, density, volume per atom. \\
  \hline
  Stoichiometric features & $L^2$ norm of fractions of elements in the material \cite{Ward2016a},  number of atoms per unit cell.\\
  \hline
  Ionic Character &  Possibility to form an ionic crystal from common oxidation states \\
  & of  the atoms \cite{Ward2018} \\
  \hline
  HOMO/LUMO  & energy and type ($s,p,d$ or $f$) of the highest occupied and lowest \\
  energies &  unoccupied atomic orbital \cite{Ward2018} \\
  \hline
  Chemical features (SAS) & covalent radius, electronegativity, electron affinity, atomic mass,  Mendeleev number \cite{Ward2016a}, melting temperature of element in pure form, atomic charge $Z^4$ (measures strength of SOC--coupling) \\
  \hline	
  Valence Orbitals (SAS) & number of unfilled $s,p,d,f$--orbitals. \\
  \hline
  Nearest--neighbor  & atomic charge, Mendeleev number, periodic table row, and column, \\
  Differences & atomic mass, covalent radius, electronegativity, Number of filled and unfilled $s,p,d,f$ valence orbitals \cite{Ward2017} \\
  \hline
  Spatial composition & bond lengths (distances to nearest neighboring atom), 
   bond angles (angles of triples of atoms), dimensionality of compound, 
   boundary area between neighbors as obtained by Voronoi tessellation \cite{Ward2018} \\
  \hline
  Wyckoff positions & multiplicity of Wyckoff position, order of the group of symmetries 
  fixing the position (stabilizer), whether the stabilizer contains inversion, 
  whether the Wyckoff position induces split EBRs  \\
  \hline
\end{tabularx}
\caption{Irrelevant, tested features by category. For all single--atom statistics (SAS) features, all nearest--neighbor difference features and the spatial composition features (except for the compound dimensionality, which is a global feature) we considered the following statistics: mean, variance, minimum, and maximum. For the Wyckoff position features we did not only consider these statistics. We also calculated a weighted average number of $s-,p-,d-$, and $f$--shell valence electrons, where each atom is weighted according to its Wyckoff position (for example, according to the multiplicity of the Wyckoff position) to capture the interaction between the symmetry of the atomic orbitals and that of an atom's position in the crystal.} 
\label{table:tested_features}
\end{table*}

In Table \ref{table:tested_features} we provide a list of the features which we tested in the course of developing our model, but which in turn proved irrelevant for the prediction of a material's topological class. In particular, we tested a number of features related to the Wyckoff positions of atoms in the crystal lattice. All the descriptors we tested failed. We believe this to be due to the effect discussed in Sect. \ref{sec:discussion}: Wyckoff position related features can give information about the topology if the wave function of a compound is similar to the non--interacting one (i.e., a wave function made out of atomic orbitals centered on the locations of the actual atoms). This is not necessarily the case for the DFT--calculated wave functions underlying the topological class given in our database. However, we cannot exclude that we simply have missed the relevant Wyckoff--related features, due to the large possible variety of such features. 

In addition to the features of Table \ref{table:tested_features} which were used together with GBT models, we also experimented with a number of machine learning models different from GBT. In particular, we tried to directly use the CGNN architecture of Ref.~\onlinecite{Xie2017}, which is available as open source software, however without success. CGNNs are similar to the convolutional neural networks which have been immensely successful in image recognition \cite{Goodfellow2016}. Crystals are represented by a graph, where each atom from the unit cell is a vertex and two atoms are joined by an edge if they are next neighbors. The vertex values undergo a series of non-linear transformations which depend only on the vertex's neighbors and are finally combined in a weighted sum, yielding the prediction. The same transformation is applied to all vertices, making it a graph convolution. Each transformation corresponds to one layer of the neural network, which is trained numerically. This procedure means that no explicit descriptor must be chosen, and the network can operate directly on crystals. However, as mentioned in Sect. \ref{sec:discussion}, the CGNN architecture expresses the quantity to be predicted as a sum of local contributions. Such an ansatz might not be appropriate for topological classes.

We also used another CGNN model from Ref.~\onlinecite{Xie2017} in an unsuccessful attempt of so--called transfer learning \cite{Pan2010}, trying to overcome the limitations due to the the size of our dataset. Transfer learning can make use of datasets for which the topological class is not known, but related quantities have been computed. The (DFT--calculated) band gap of a material provides useful information on a material's topology (for instance, semi--metals are gapless and large gaps of more that 1eV cannot be topological). Indeed, the TI--$F_1$--score of our model improves from 70 to 75\% if we provide the \emph{actual} DFT gap. The idea was to predict the band gaps with the CGNN and then feed this prediction as an additional feature to our GBT model. While this particular attempt did not work (likely because the predicted gaps were not sufficiently accurate), the general strategy, which is related to the ML concept of \textit{stacking} \cite{Rokach2010}, seems promising. In particular, it can make use of datasets for which some relevant quantities like the band gap have been computed, but the topological class has not been computed and combines different descriptors which would be unsuccessful on their own.
  
Further, we considered some of the descriptors previously discussed in the literature \cite{MatthiasRupp2011,Bartok2013,K.T.Schuett2014,
KatjaHansen2015,MatthewHirn2017,Huo2018,
GabrielR.Schlede2019}. A large number of these are not well adapted to samples containing a wide range of chemical elements and many atoms per unit cell because the dimension $d$ of the descriptor becomes very large. We considered, for example, the sine matrix descriptor \cite{FelixFaber2015} in combination with a kernel support vector machine classifier \cite{Scholkopf} (with Gaussian kernel), an approach common in the literature \cite{FelixA.Faber2017}. For a material with $N_\text{atom}$ atoms with nuclear charges $\{Z_i\}_{i=1,...,N_\text{atom}}$ at positions $\{\mathbf{R_i}\}_{i=1,...,N_\text{atom}}$ in the unit cell it is the $N_\text{atom}\times N_\text{atom}$ matrix defined by
\begin{align}
    C_{ij}^\mathrm{sine}=\begin{cases}
    \frac{Z_i Z_j}{\lvert \mathbf{B} \cdot \sum_{k=\{x,y,z\}} \hat{\mathbf{e}}_k \sin^2 \left( \pi \mathbf{B}^{-1} \cdot \left( \mathbf{R}_{i} - \mathbf{R}_{j} \right) \right)\rvert} & \text{for } i \neq j \\
    \frac{Z_i^{2.4}}{2} & \text{for } i = j 
    \end{cases}.
\label{eq:sine}
\end{align}
Here, $\mathbf{B}$ is the $3\times 3$ matrix formed from the lattice basis vectors. Empirically, we find that such a model cannot distinguish different topological classes. The sine matrix was originally conceived as a proxy for the electrostatic force field due to the atomic nuclei. This makes it suitable, for instance, for models which predict mechanical properties like the bulk modulus. In contrast to the former, topological classes have no clear link to interatomic force fields and the failure of this and related descriptors like the Ewald matrix \cite{FelixFaber2015} is unsurprising.

Finally, approximately 80\% of the materials in the ICSD are classified into so--called material prototypes according to their crystal structure \cite{ICSD}. For example, the NaCl--type is defined to contain all materials from SG 225 with two different atoms per unit cell at positions $[0,0,0]$ and $[\tfrac{1}{2},\tfrac{1}{2},\tfrac{1}{2}]$ in the standard basis of lattice vectors (this is the crystal structure of NaCl). We selected the 100 most common material prototypes (a total of 10389 materials) and encoded the prototype with the one--hot method as defined by Eq. \ref{eq:one_hot}. This approach also failed to improve over the model shown in Table \ref{table:performance}.

\section{\label{methods:categorical} Encoding of Categorical Features}

The global features SG and $N_e$ we described are \textit{categorical}: they indicate membership in a class, for example in the materials with a certain SG. This fact should be reflected by the way the feature is encoded as a real vector for the ML algorithm \cite{foundations}: Most algorithms assume that two samples are similar if the difference of the two corresponding vectors is small. However, e.g. SG 9 contains no crystalline TIs whereas in SG 10 47\% of materials are crystalline TIs. Accordingly, the actual number of a SG is not a reasonable encoding by itself -- this number is somewhat arbitrary (but not completely: SGs are ordered according to point group and lattice type). We therefore supplement the SG number with an additional descriptor. We calculate the following frequencies for each SG $g\in\{1,...,230\}$ and topological class $i\in\{\text{ Trivial}, \text{ NLC}, \text{ SEBR}, \text{ ES},\text{ ESFD}\}$:
\begin{align}
\frac{\text{Number of materials with top. class } i \text{ and SG } g}{\text{Number of materials with SG } g}
\label{eq:SG_frequencies}
\end{align}
This can be interpreted as the conditional probabilities to belong to a topological class given the SG. Eq. \ref{eq:SG_frequencies} is then used to encode the SG, e.g. a material with SG 10 would be assigned the vector [0.42 (trivial), 0.47 (NLC), 0 (SEBR), 0 (ES), 0.11 (ESFD), 10 (SG number)].

In particular, when we evaluate a model using a training and a test set, Eq. \ref{eq:SG_frequencies} is calculated using the materials in the training set only. Else, information about the test set (namely, how topological class and SG are related in the test set) would indirectly be available when the model is trained (this is called target leakage in ML). Note that using \textit{only} Eq. \ref{eq:SG_frequencies} and not the raw SG number only lowers the TI--$F_1$ score by approximately 1\% point.

For similar reasons, $N_e$ is not fed directly into the model but instead represented by the digits in its binary representation, e.g. $N_e=7$ would be represented by the vector $[1,1,1]$. This makes it easy for a decision tree to check for example whether $N_e$ is odd. Recall that the leaf of a decision tree corresponds to a connected region in the space of the input vectors (see Fig. \ref{figure:decision_tree}). So a tree which splits the numbers $1,...,m$ into even and odd numbers based on their numerical value needs to split $[1,m]$ into $m$ connected regions (here, intervals) and thus has $m$ leaves. With the binary encoding, a tree with two leaves suffices.

We also tested different ways of encoding the SG. One often--used option to encode categorical features is \textit{one--hot} encoding. For the case of $N_c$ different classes, it is defined by the following mapping:
\begin{align}
\{1,...,N_c\} \rightarrow \mathbb{R}^{N_c}, \quad i \mapsto \mathbf{e}_i
\label{eq:one_hot}
\end{align}
where $\mathbf{e}_i$ is the $i^\text{th}$ unit vector. However, the method described above proved superior.

\section{\label{methods:random_classifier} Random Classifiers}

The performance of a classifier depends strongly on the frequency of the different classes in the dataset. Let us consider a binary classifier and a dataset where the fraction of positive samples is $q$. To understand the dependence of a classifier on $q$, it is useful to consider a toy example: the random classifier, which assigns a positive label with probability $p$ and a negative one with probability $1-p$ (a purely positive classifier is the special case $p=1$). Now the dependence of various metrics on $q$ is simple to calculate.  For the metrics considered in this paper, one has:
\begin{align}
&\text{accuracy}=pq+(1-p)(1-q), \\
&\text{precision}=q, \; \text{recall}=p, \; F_1=2\frac{pq}{p+q}.
\label{eq:random_classifier}
\end{align}

\begin{figure*}[t]
	\centering
   \includegraphics[scale=0.3]{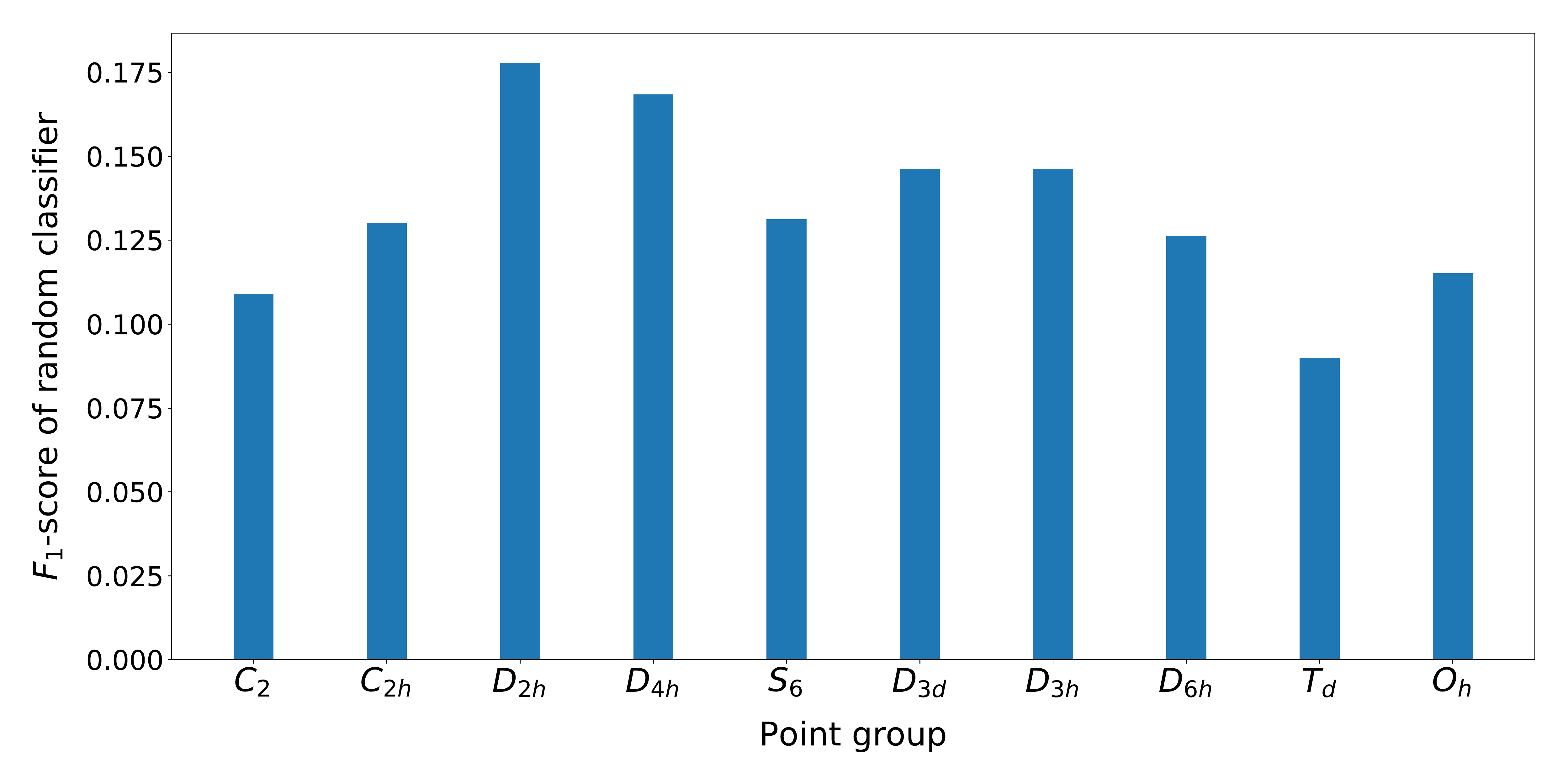}
   \caption{Dependence of the $F_1$--score of a random classifier on the the PG. Each PG contains a different fraction $q$ of TIs, so according to Eq. \ref{eq:random_classifier} a random classifier will perform differently on them. While these $F_1$--scores are of course much lower than those of Fig. \ref{figure:point_groups} (please note that the vertical scale is different), they can explain the variation of the actual scores between the first seven PGs ($\bar{1}$ to $\bar{6}2m$). However, the performance drop observed in Fig. \ref{figure:point_groups} for the last three PGs is much more severe than would be expected from the random classifier model.}
   \label{figure:random_F1_PG}
\end{figure*}

These results can help explain patterns seen in real classifiers and in particular highlight behavior which deviates from these simple models. First, we can judge the quality of our model by comparing it to a random classifier, as done in the precision--recall curve of Fig. \ref{figure:precision_recall_curve}. Second, the $F_1$--score of the random classifier can partially explain the behavior seen in Fig. \ref{figure:point_groups}. Here, the frequency of  positive samples, i.e., TIs, varies between different datasets, i.e., materials from different point groups, while the classifier remains the same. We therefore consider a random classifier with $p$ fixed to 0.14 (the frequency of TIs in the overall dataset). The results are shown in Fig. \ref{figure:random_F1_PG}.

\section{\label{methods:nice} High-Confidence Topological Materials}

As discussed in Sect.~\ref{sec:limitations}, we believe that a part of the errors of our model is due to unreliability of the DFT calculations. To test this hypothesis, we selected a of number topological materials (according to the criteria detailed below) where we have a high confidence in the predictions of DFT. We refer to these compounds as \textit{high-confidence} topological materials or HC materials. 

We then tested the performance of our model exclusively on HC materials, using a modified cross--validation scheme. We split the HC materials into five stratified folds (see Appendix \ref{methods:cross_val}), trained the model on the union of  the non--HC materials and four of the folds, and finally evaluated the model on the remaining fold (the results are displayed in Table \ref{table:subclasses_nice}). This way, we ensure that some HC materials are present in the training set. Otherwise, training and test set would contain very different materials, which makes it impossible to make an estimate of the model performance. 

To select the HC materials, we used criteria based on both properties of the materials and of the DFT results. Regarding the first, we exclude $f$--electron and magnetic materials. Regarding the latter we chose TIs with large direct gaps ($>0.025$eV for NLC and $>0.035$eV for SEBR, corresponding to the highest 20\% of gaps) and TSMs with low numbers ($>32$ for ES and $>25$ for ES) of Fermi level crossings. This way, we  selected a total of 1978 rivial, 249 NLC, 316 SEBR,  404 ES and 1090 ESFD materials. The proportions of the topological classes among the HC materials are almost the same as in the entire dataset, to avoid any distortions when testing a model on the HC materials.


\end{document}